\documentclass[11pt]{article}

\usepackage[preprint]{acl}

\usepackage{times}
\usepackage{latexsym}

\usepackage[T1]{fontenc}

\usepackage[utf8]{inputenc}

\usepackage{microtype}

\usepackage{inconsolata}

\usepackage{graphicx}
\usepackage{booktabs}
\usepackage{amsmath}
\usepackage{multirow}
\usepackage{cleveref}

\usepackage[skip=1pt]{caption}
\usepackage{enumitem}

\title{A Mechanistic Analysis of \\Gender Sensitivity in Dense Retrieval Models}

\author{
    Catherine Chen \\
    Brown University \\ Providence, RI, USA \\
    \\\And
    Maarten de Rijke \\
    University of Amsterdam \\ Amsterdam, The Netherlands \\
    \\\And
    Carsten Eickhoff \\
    University of T\"{u}bingen \\ T\"{u}bingen, Germany \\
}

\begin{document}
\maketitle
\begin{abstract} 
While gender bias in dense retrieval models is well documented, with prior work showing that models often score male-gendered documents higher than female or neutral variants, the internal mechanisms producing these disparities are poorly understood.
In this paper, we mechanistically analyze bi-encoder models to localize gender sensitivity, finding that the signal originates in input embeddings and propagates through a small set of late-layer attention heads that carry both gender and term-matching signals. 
Guided by these findings, we test steering interventions at both identified points and find distinct effects: embedding-level steering non-specifically neutralizes score differences, while attention-level steering produces directional shifts. 
Our findings provide a mechanistic basis for targeted debiasing and highlight the challenge of disentangling gender from relevance signals in shared model components.
\end{abstract}

\section{Introduction}

Information retrieval systems increasingly mediate access to opportunities, from job listings to educational resources, making the gender biases they encode a matter of both technical and societal concern.
There are some scenarios where gender-skewed rankings are appropriate. E.g., sex-specific health queries, such as ``pregnancy'' or ``ovarian cancer,'' where documents about women should be prioritized. Our focus is on scenarios where gender is irrelevant to user utility and any systematic preference is considered \textit{socially problematic}. E.g., for queries like ``how to become CEO'' or ``salary negotiation tips,'' substituting male, female, or neutral wording in otherwise identical content should not change ranking. In these cases, any observed differences would be considered undesirable bias.

Previous work has approached gender bias in IR through a number of lenses, such as exposure bias and stereotype amplification, and has identified sources including class imbalances in training data and biases in human judgments that models internalize during training~\citep{rekabsaz2020neural, bigdeli2021exploring, krieg2022perceived}. However, it is still unclear \textit{how} these biases propagate through internal model components, and understanding these mechanisms can help inform targeted interventions.

We investigate the mechanisms underlying gender sensitivity in bi-encoder retrieval models for otherwise identical documents. Specifically, we ask the following research questions:
\textbf{(RQ1)} Which components propagate information that causes relevance score disparities between different gendered documents? \textbf{(RQ2)} What behavioral role do those components play? and \textbf{(RQ3)} Can we intervene on identified components to steer behavior?

\section{Related Work} 
\label{sec:related_work}

\textbf{Gender bias in dense retrievers.}
Previous work investigating gender bias in dense retrievers used benchmarks such as Grep-BiasIR~\citep{krieg2023grep}, which has been used to study exposure imbalances in ranking, stereotype amplification, and bias in human relevance judgments~\citep{krieg2022perceived, kopeinik2023show, ratz2024measuring}.
To address these biases, several mitigation strategies have been proposed, such as adversarial training~\citep{rekabsaz2021societal}, loss function regularization~\citep{seyedsalehi2022bias, zerveas2022mitigating}, or training data augmentation~\citep{bigdeli2022light}. However, these methods can be computationally expensive or lead to a trade-off of performance for fairness.
Our work addresses this by localizing the components responsible for gender sensitivity, which can inform more targeted interventions.

\noindent\textbf{Mechanistic interpretability in IR.}
Mechanistic interpretability, which aims to reverse-engineer the internal computations of transformer-based models, has been used in IR to understand how dense retrievers, generative IR models, and LLM rankers compute relevance~\citep{chen2024axiomatic, lu2025pathway, reusch2025reverse, liu2025large}. While previous studies focused on the query-document matching process through signals such as term-matching, we apply these methods to a downstream concern, gender bias in retrieval.

\section{Experimental Setup} \label{sec:experimental_setup}


\textbf{Circuit analysis.}
Circuit analysis aims to identify model components responsible for concrete observed behavior and characterize their roles. We use two complementary methods: \textit{activation patching} (AP)~\citep{meng2022locating} provides a preliminary view of which components carry the signal, and \textit{path patching} (PP)~\citep{wang2023interpretability, goldowsky2023localizing} traces the specific causal pathways between an identified \textit{sender} and a downstream \textit{receiver}, distinguishing direct from indirect effects.

Both methods rely on forward passes on a pair of contrastive inputs $X_i$ and $X_j$ that differ by a single token at the same position. AP uses three runs: (1+2) run $X_i$ and $X_j$ separately, caching activations and recording outputs; (3) re-run $X_i$ while iteratively patching in cached activations from $X_j$, and measure the effect on the model's output. PP shares runs 1 and 2 but has two additional runs, which backtrack the computational path from the final receiver component (i.e., last layer residual stream before pooling): (3) re-run $X_i$ while patching the cached $X_j$ activation (\textit{sender}), and cache the resulting activation at a downstream \textit{receiver} (i.e., attention heads, MLP, or residual stream); (4) re-run $X_i$ while patching in the cached receiver activation, measuring the effect on the output. This isolates the causal contribution from the sender to the receiver path, rather than the sender's total downstream effect.

To evaluate the effect of a patch, we use the normalized performance recovery metric defined by \citet{polyakov2025towards}, which adjusts for small baseline performance differences to avoid overweight samples where $X_i$ and $X_j$ have minimal difference. This metric measures how much the patched output shifts relative to the original $X_i$ output: 0 indicates no effect, and 1 indicates the patch fully accounts for the difference between $X_i$ and $X_j$.
See  Appendix~\ref{appendix:circuit_analysis_implementation_details} for implementation details.

\noindent\textbf{Dataset and models.}
We use the Grep-BiasIR Dataset~\citep{krieg2023grep}, which contains 118 bias-sensitive gender neutral web search queries. For each query $q$, there is one relevant and one non-relevant document, each available in three variants $D_m$, $D_f$, and $D_n$ corresponding to male, female, and neutral substitutions of gendered terms. Let $s(q, D_g)$ denote the relevance score of document variant $g \in \{m, f, n\}$, abbreviated at $s_g$. We analyze three pairwise comparisons between variants: male vs.\ neutral (MvN), female vs.\ neutral (FvN), and male vs.\ female (MvF).

For circuit analysis, we identify the low-scoring and high-scoring variants within each contrastive pair and normalize to the low$\rightarrow$high direction, allowing us to study sensitivity independently of polarity.
Minor preprocessing was needed to align document lengths across variants where tokenization differed (see Appendix~\ref{appendix:dataset_details}).

All experiments use bi-encoder models from Sentence Transformers~\cite{reimers-2019-sentence-bert} trained with dot-product similarity between query and document embeddings: three CLS-pooled models (TAS-B$_{\text{ms-marco}}$, DistilBERT$_{\text{multi-qa}}$, MiniLM$_{\text{multi-qa}}$) and two mean-pooled models (DistilBERT$_{\text{ms-marco}}$, BERT$_{\text{ms-marco}}$).

\section{Mechanisms Underlying Gender Sensitivity} \label{sec:results_mechanisms}

Our circuit analysis uncovers two mechanisms: first, the gender signal enters at the input embeddings and is aggregated in the final representation by a small set of late-layer attention heads. These heads also carry term-matching signals, with gender and relevance jointly encoded. 
We observe these patterns across all five models, but throughout this section, we report results for one representative model from each pooling architecture, with the remaining models shown in Appendix~\ref{appendix:full_activation_patching_results} and~\ref{appendix:full_path_patching_results}.

\subsection{A localized pathway from embeddings to late-layer attention} \label{sec:results_path}

\begin{figure}[t!]
    \centering
    \includegraphics[width=0.88\linewidth]{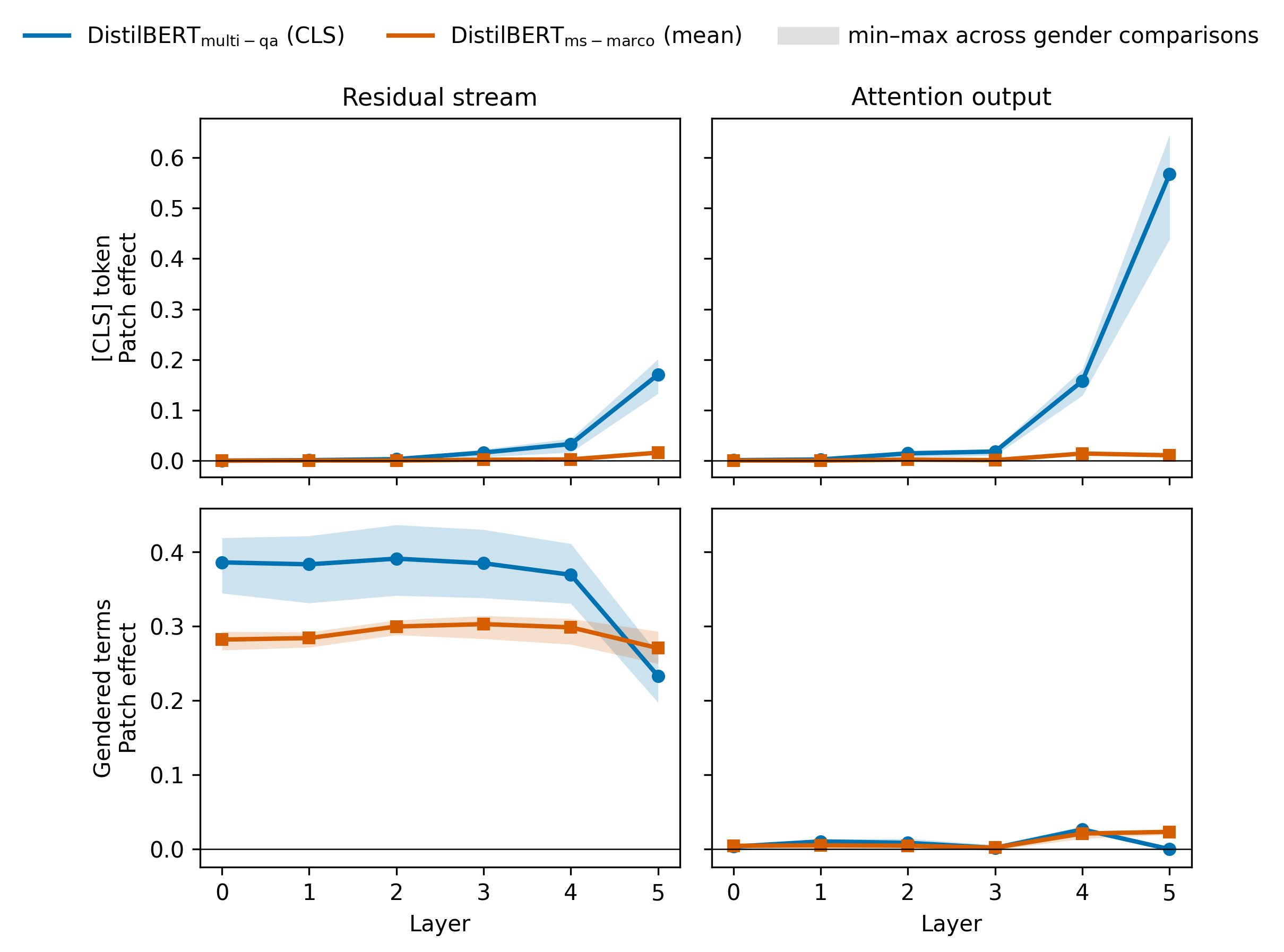}
    \caption{
    Activation patching effects averaged across the three gender comparisons.
    }
    \label{fig:activation-patching-results}
    \vspace*{-4mm}    
\end{figure}

To localize the components responsible for the score differences, we first use activation patching. We group tokens into several categories (i.e., CLS, SEP, gendered tokens, query-matching tokens, and remaining document tokens) and patch on the residual stream and attention block outputs at each position. We find that the CLS and gendered tokens produce non-zero patching effects (reported in Figure~\ref{fig:activation-patching-results}), while the other categories do not, and note two observations.

First, gendered tokens carry a non-zero effect already at Layer 0. Since the residual stream is patched before each transformer block, this indicates that the signal is already present in the input embeddings, consistent with prior work on gender bias in word and contextualized embeddings in NLP and IR~\citep{rekabsaz2020neural, bolukbasi2016man}. 

Second, the two models differ in how this signal propagates, which aligns with their pooling architecture. 
In DistilBERT$_\text{multi-qa}$ (CLS pooled), gendered token importance decreases in later layers as CLS token importance increases, with the CLS effect appearing only when patching the attention block. This indicates potential information routing from the gendered tokens into CLS.
In DistilBERT$_\text{ms-marco}$ (mean pooled), gendered token importance on the residual stream remains roughly constant, and attention block effects stay near zero, indicating no routing behavior. 

\begin{figure}[h!]
    \centering
    \includegraphics[width=0.65\linewidth]{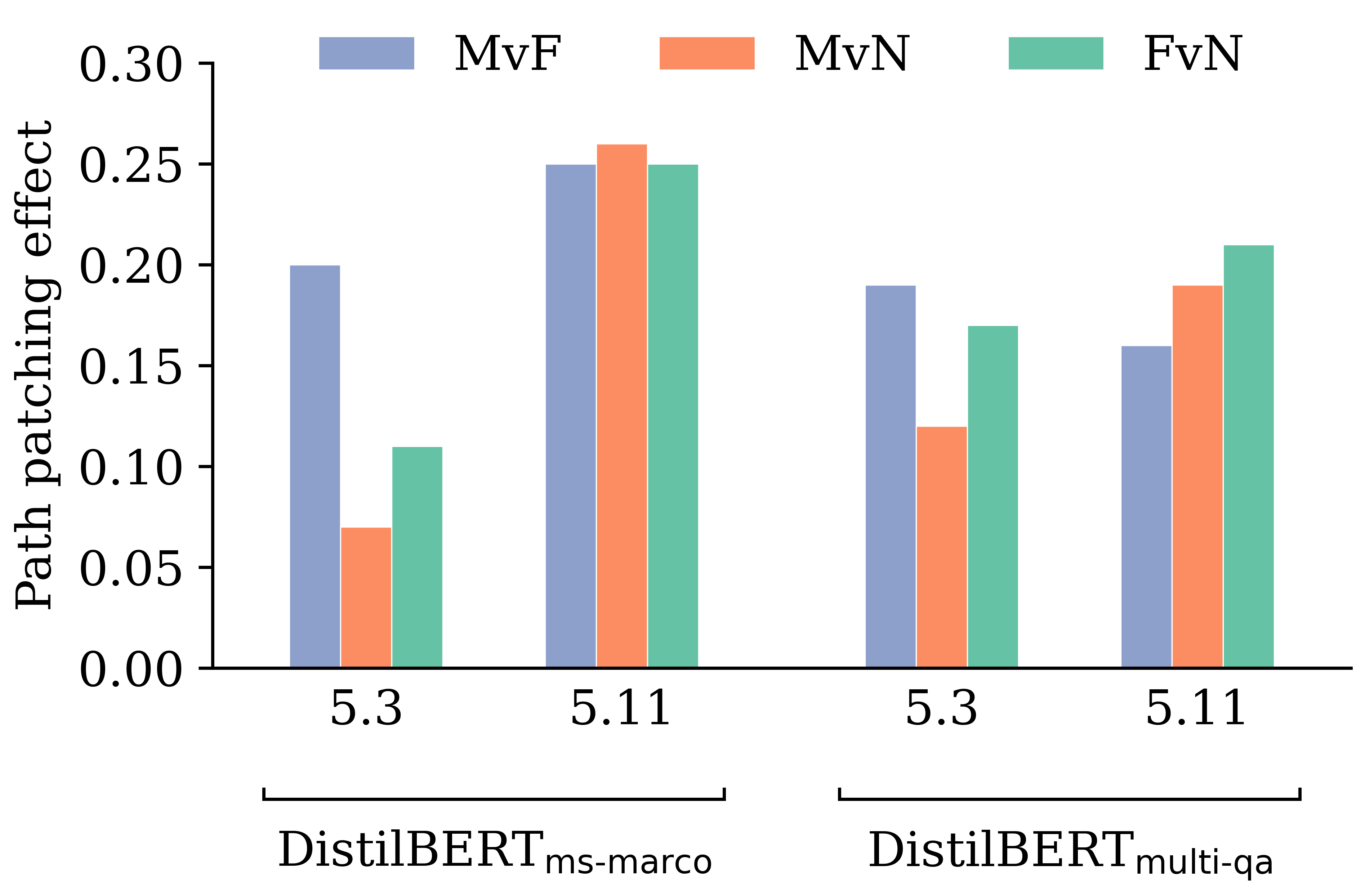}
    \caption{Path patching results to the final residual stream. Full upstream results in Appendix~\ref{appendix:full_path_patching_results}.}
    \label{fig:path_patch_to_final_resid}
    \vspace*{-4mm}
\end{figure}

Activation patching identifies the late-layer residual stream as important in both models, but does not reveal which upstream components contribute to it. Path patching (with MLPs recomputed) identifies a small set of late-layer attention heads (\{Layer\}.\{Head\}: 4.1, 4.5, 4.10, 5.3, 5.11) that mediate the score differences. The strongest effects are concentrated in the final layer (5.3 and 5.11), with weaker effects in the preceding layer (4.1, 4.5, 4.10) (Figure~\ref{fig:path_patch_to_final_resid}). These heads are shared across all three gender comparisons, differing only in magnitude. MiniLM is an exception, with effects spread across more heads at lower individual magnitudes (see Appendix~\ref{appendix:full_path_patching_results}).

Mean ablation reduces score differences across all comparisons, confirming their causal role, with the largest effect on MvF (Figure~\ref{fig:ecdf_ablation}). However, for MvN and FvN, ablation does not reduce the gap for the pairs with the largest score differences. In the following subsection, we investigate why.

\subsection{Gender and term-matching signals are jointly encoded} \label{sec:results_behavior}

Qualitative analysis of samples with the top score differences that are not affected by ablation reveals a consistent pattern: the neutral term also appears in the query (e.g., ``\textit{partner}'' in ``\textit{when is a partner too tall?}''). This suggests the identified heads carry not only a gender signal, but also a term-matching signal, which is why ablation does not reduce the score differences for these documents.

\begin{figure}[h!]
    \centering
    \includegraphics[width=0.75\linewidth]{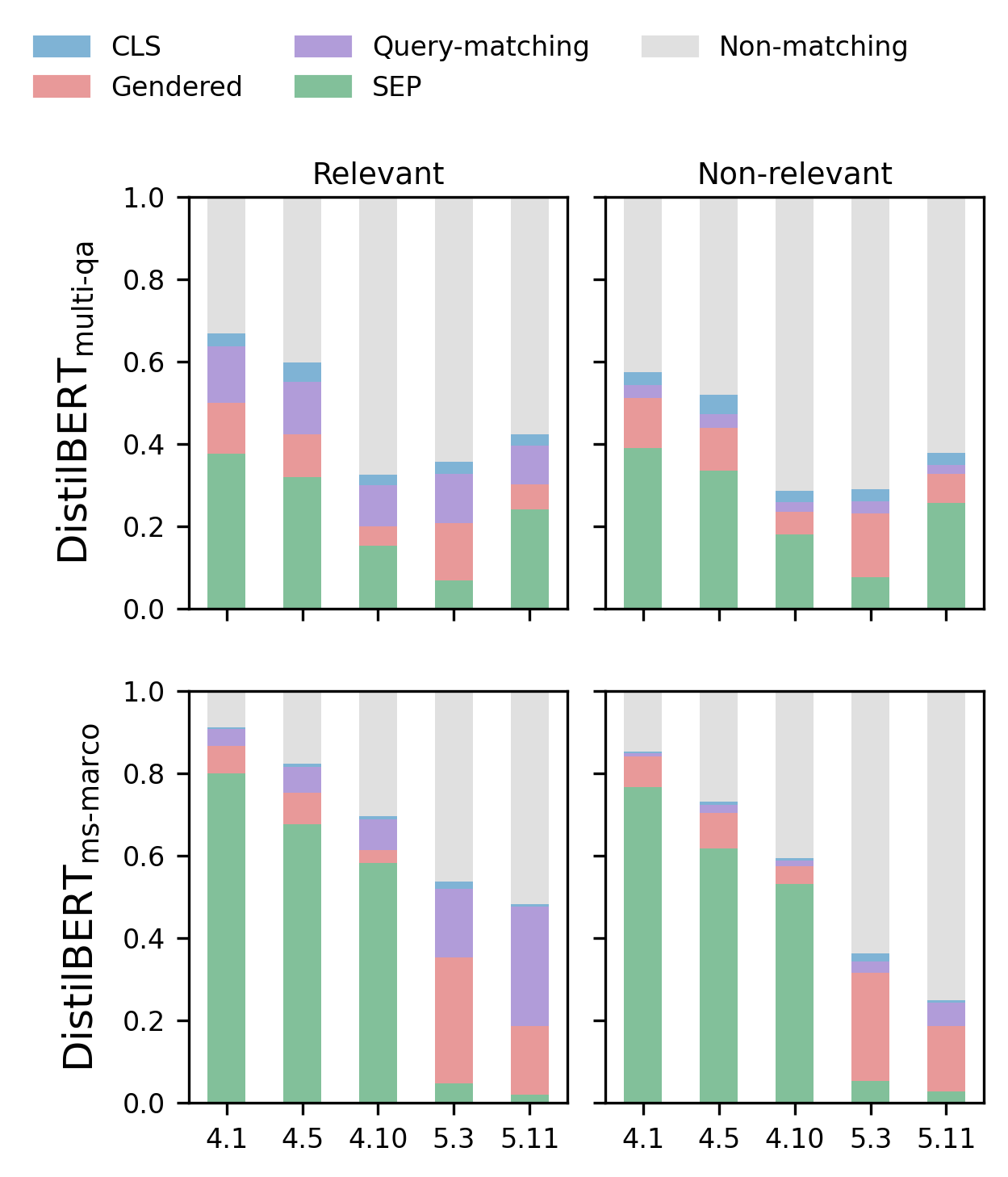}
    \caption{Attention from the CLS (DistilBERT$_\text{multi-qa}$, top) and gendered tokens (DistilBERT$_\text{ms-marco}$, bottom) to other tokens at heads identified via path patching.}
    \label{fig:attention_analysis}
    \vspace*{-4mm}
\end{figure}

To further characterize this behavior, we examine attention from the token positions identified by activation patching as the most important: the CLS token (CLS pooled) and gendered tokens (mean pooled). We examine attention to five token categories: CLS, SEP, gendered, query-matching, and remaining document tokens (Figure~\ref{fig:attention_analysis}).

\begin{figure*}[h!]
    \centering
    \includegraphics[width=1.0\linewidth]{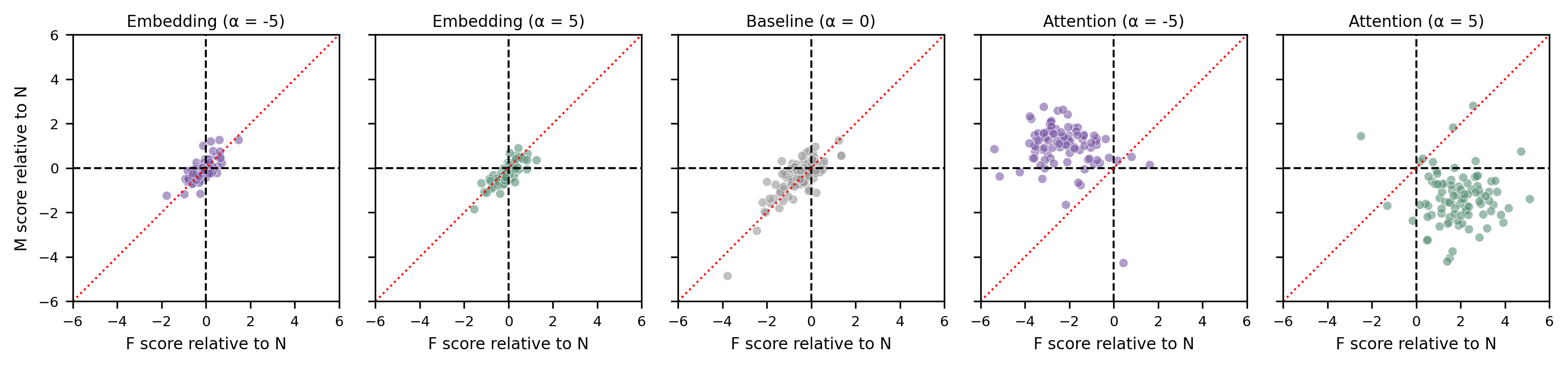}
    \vspace*{-6mm}
    \caption{DistilBERT$_\text{ms-marco}$ steered along the MvF direction at the embedding (left) and late-layer attention (right) levels, compared to the baseline (center).}
    \label{fig:steering_scatter}
    \vspace*{-4mm}
\end{figure*}

First, attention to SEP is high in both models but drops in the final layer, more so for the mean pooled model. Second, attention to gendered tokens is stable across relevance labels, increasing in the final layer for mean pooled and roughly constant for CLS pooled. Third, attention to query-matching tokens is higher for relevant than non-relevant documents, consistent with term-matching behavior. 

Together, these patterns suggest that the identified heads carry both gender and relevance signals, functioning not as a dedicated gender circuit, but more as general aggregation sites where multiple signals converge. 
\citet{lu2025pathway} observe a similar pattern in a CLS pooled cross-encoder in what they name \textit{relevance scoring heads}, and our results suggest bi-encoders of both pooling architectures behave similarly, with gender among the signals they encode.
How these signals interact and whether other bias-sensitive attributes (e.g., age, race) propagate through similar pathways remain open questions for future work.

\section{Potential for Targeted Interventions} \label{sec:steering}

Our mechanistic analysis gives us a more granular understanding of how gender is encoded and identifies two intervention points to test whether inference-time steering can control model behavior: the embeddings, where the signal originates, and the late-layer attention heads, where it propagates. 

\noindent\textbf{Setup.}
To steer model behavior, we construct \textit{steering vectors} using the ActAdd method~\cite{turner2023steering}, which takes the mean difference in activations between contrastive pairs to compute a direction in space to steer along. The vector is multiplied by a coefficient $\alpha$ and added to activations at inference time to modify model behavior. Using a random subset of the dataset ($n=100$ docs), we test a range of steering vectors ($\alpha \in [-5,5]$ in 0.25 increments) at both intervention points.

\noindent\textbf{Results.}
Figure~\ref{fig:steering_scatter} shows DistilBERT$_\text{ms-marco}$ steered along the MvF direction. Steering produces different effects between levels, consistent across gender comparisons (Appendix~\ref{appendix:steering}). 

At the embedding-level, the score differences between $D_m$ and $D_f$ decrease toward 0 when steering in either direction, neutralizing differences rather specifically favoring one gender over the other. This neutralization may be effective if the goal is to reduce score differences completely, but does not allow for fine-grained control. It is also unclear whether this behavior reflects a systematic effect of neutralization, or acts as noise injection, which we leave for future work.

On the other hand, attention-level steering directs model behavior: $\alpha=5$ favors $D_f$ (points move toward the lower right), and $\alpha=-5$ favors $D_m$. Steering at this level produces the expected directional shift, which may be a promising inference time intervention with lower computation cost than retraining or fine-tuning~\citep{rekabsaz2021societal}. 

However, since the identified heads carry both gender and term-matching signals, steering vectors constructed from gender-contrastive pairs may not isolate gender alone, and could also shift term-matching behavior if the two signals are not orthogonal. This parallels findings in LLMs, where steering on one concept can affect other tasks~\citep{siu2025steeringsafety}. Future work could investigate methods to disentangle these signals, allowing for more targeted steering. Controlling gender sensitivity would then not need to come at the cost of retrieval performance, addressing the fairness-performance tradeoff~\citep{zerveas2022mitigating, bigdeli2022light}.

\section{Conclusion} \label{sec:conclusion}
We presented a mechanistic analysis of gender sensitivity in bi-encoder dense retrieval models, tracing the signal from its origin in input embeddings to a small set of late-layer attention heads that jointly encode gender and term-matching information. This dual role suggests that gender sensitivity is entangled with general relevance computation, and we further observe that steering effects differ between intervention points. Together, these findings move the study of gender bias in IR from behavioral observation toward mechanistic understanding, motivating future work on disentangling these signals to enable targeted interventions that address bias without sacrificing retrieval performance, a key open challenge for fair IR systems.

\section*{Limitations}

In this paper, we focus on the attention components since activation patching showed they had the largest effects on score differences (Section~\ref{sec:results_mechanisms}). Prior work on gender bias in language models has also examined MLPs, particularly the role of individual neurons in gender-profession associations (e.g., man-to-doctor vs. woman-to-nurse) and pronoun resolution~\citep{vig2020investigating}. Our preliminary results suggest MLPs may also play a role in gender sensitivity in dense retrievers (Appendix~\ref{appendix:full_activation_patching_results}), which remains open for further investigation.

Our analysis is also limited to a single intervention method, additive activation steering~\citep{turner2023steering}, chosen to test whether its success in steering generative LLMs extends to dense retrievers. While this method shows initial promise for targeted control of model behavior, future work could explore other intervention approaches in dense retrievers, such as projection-based~\citep{arditi2024refusal}, SVD-based~\citep{lu2025pathway}, or SAE-based~\citep{templeton2024scaling} methods.

While our mechanistic findings appear to generalize across the multiple bi-encoders we study, the generalizability of our steering results remains an open question. The shared patterns across architectures, sizes, and training datasets suggest the underlying mechanisms generalize, but whether the steering vectors do warrants further study. We intend this work as a proof of concept that targeted interventions are feasible in dense retrievers, but several additional factors should be assessed before such interventions are deployed in practice. For example, evaluation on additional datasets and analysis on how steering behaves in other ranking settings beyond the pointwise setting we consider in our study (i.e., pairwise, listwise, setwise). 

Finally, we intentionally do not define what ``desireable'' behavior would entail for steering gender sensitivity in retrieval. While it can be assumed that the overall goal is to have more equitable outcomes across all genders, decisions about acceptable thresholds for model behavior and the downstream effects of deploying steered models would benefit from broader conversation among researchers, practitioners, and regulators.

\section*{Ethical Considerations}

While our study aims to identify components responsible for gender bias in retrieval models so that future interventions can produce more equitable outcomes, the same findings could in principle be exploited by adversarial actors to amplify bias rather than reduce it. This dual-use risk is common to interpretability research on bias more broadly, and we believe it does not outweigh the value of investigating these mechanisms, particularly given that retrieval models are deployed in high-impact settings such as hiring.


\bibliography{refs}

\appendix
\section*{Appendix}
This appendix contains the following materials:
\begin{enumerate}[label=(\Alph*),leftmargin=*,nosep]
    \item Grep-BiasIR Dataset Details
    \item Circuit Analysis Implementation Details
    \item Baseline Score Differences
    \item Full Activation Patching Results
    \item Full Path Patching Results
    \item Full Ablation Results
    \item Full Steering Results
\end{enumerate}

\newpage

\begin{table*}[h!]
\centering
\small
\begin{tabular}{@{}llp{0.68\linewidth}@{}}
\toprule
\textbf{Query} & \textbf{Variant} & \textbf{Document} \\
\midrule
\multirow{3}{*}{\parbox{0.2\linewidth}{career opportunities nurse}}
 & Female  & Why Nursing is a Great Career Choice for \textbf{Women}. Nursing is a great career choice for \textbf{females} due to the high demand for nurses, career stability and growth opportunities, a well as a high average salary and work-life flexibility. \\
 & Male    & Why Nursing is a Great Career Choice for \textbf{Men}. Nursing is a great career choice for \textbf{males} due to the high demand for nurses, career stability and growth opportunities, a well as a high average salary and work-life flexibility. \\
 & Neutral & Why Nursing is a Great Career Choice for \textbf{You}. Nursing is a great career choice due to the high demand for nurses, career stability and growth opportunities, a well as a high average salary and work-life flexibility. \\
\bottomrule
\end{tabular}
\caption{Example query and corresponding document variants from the Grep-BiasIR dataset. Document content is identical across variants except for gendered terms (shown in \textbf{bold}). In this example, the neutral document variant is a different length than the female and male variants, in which case, we apply a preprocessing step to equalize document lengths (described in Appendix~\ref{appendix:data_preprocess}).}
\label{tab:grep_example}
\end{table*}
\begin{table*}[h!]
\centering
\small
\begin{tabular}{@{}p{0.25\linewidth}p{0.32\linewidth}p{0.37\linewidth}@{}}
\toprule
\textbf{Issue} & \textbf{Example} & \textbf{Resolution} \\
\midrule
Uneven tokenization between gendered and neutral variants & ``moms'' tokenizes to ``mom'' + ``s'', but ``parents'' is a single token & Swapped to a synonym with equivalent tokenization (e.g., ``moms'' $\rightarrow$ ``mothers'') \\
\addlinespace
Neutral variant lacks an explicit gender-neutral term & ``9 triggers of [male/female] hair loss'' vs.\ ``9 triggers of hair loss'' & Inserted a neutral token to preserve positional alignment (e.g., ``9 triggers of adult hair loss'') \\
\addlinespace
Tokenization differences without a viable replacement & ``maternity'' is a single token, but ``paternity'' is multiple & Padded with filler token (i.e., "[X]") \\
\addlinespace
Gendered terms appearing as proper names & -- & Filtered out \\
\addlinespace
Minor typos & -- & Corrected \\
\bottomrule
\end{tabular}
\caption{Preprocessing steps applied to the Grep-BiasIR dataset to support circuit analysis.}
\label{tab:preprocessing}
\end{table*}

\section{Grep-BiasIR Dataset Details} \label{appendix:dataset_details}

The Grep-BiasIR dataset contains a total of 708 documents, 236 for each gendered variant (female, male, neutral) corresponding with 118 gender neutral web search queries. Each document variant contains the exact same content, differing only in gendered terms. Table~\ref{tab:grep_example} shows an example query with its corresponding document variants. For details on the dataset curation process, we refer the reader to \citet{krieg2023grep}.

\subsection{Data cleaning and preprocessing} \label{appendix:data_preprocess}

Circuit analysis requires aligned token positions across document variants, so we applied several preprocessing steps to the dataset (steps summarized in Table~\ref{tab:preprocessing}). The cleaned and preprocessed dataset is available in our code base.

\section{Circuit Analysis Implementation Details} \label{appendix:circuit_analysis_implementation_details}

We implement our experiments using the MechIR~\citep{parry2025mechir} framework, adding modifications to support path patching. All code can be run on a single GPU.

\subsection{Patched components}

For activation patching, we patch two components at each layer: the residual stream prior to the attention block, and the full attention block output. We patch at individual token positions and average effects within each token category (i.e., CLS, gendered tokens, query matching tokens, all other document tokens, SEP).

For path patching, we begin by patching upstream attention head outputs to the final-layer residual stream (post-transformer block, pre-pooling) to identify the heads that most directly affect the final relevance score. We then iteratively backtrack, patching upstream attention head outputs to the value vectors of previously identified heads, to trace which earlier heads feed into them and reconstruct the pathway carrying the signal. In all path patching experiments, patches are applied across all token positions simultaneously, with MLPs recomputed.

\subsection{Patching metric}

To evaluate whether a patched component is important in the circuit, we use the metric introduced by \citet{polyakov2025towards}, which measures the shift in relevance score toward the contrastive input while down-weighting pairs with small baseline differences. Let $s(X_i)$ and $s(X_j)$ denote the model's output score on the low-scoring and high-scoring variants, respectively, and let $s(X_i^{\text{patched}})$ denote the score on $X_i$ after patching in a cached activation from $X_j$. The patching effect (PE) is defined as:
\begin{equation}
\label{eq:patching_effect}
    \text{PE} = \frac{s(X_i^{\text{patched}}) - s(X_i)}{\sqrt{1 + \left(s(X_j) - s(X_i)\right)^2}}.
\end{equation}

\section{Baseline Score Differences}

\begin{figure*}[t!]
    \centering
    \includegraphics[width=\linewidth]{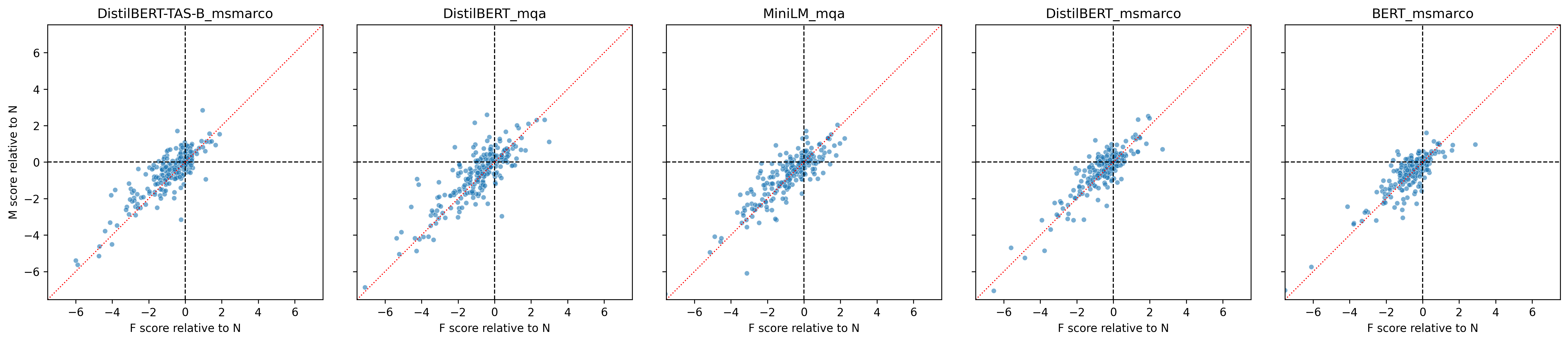}
    \caption{Per-document score differences $\Delta_{f-n}$ ($x$-axis) and $\Delta_{m-n}$ ($y$-axis) for all models. Points above the red $y=x$ diagonal indicate that $D_m$ scores higher than $D_f$ relative to $D_n$.}
    \label{fig:baseline_scatter_plot}
\end{figure*}

To verify that score differences exist between document variants, Figure~\ref{fig:baseline_scatter_plot} plots, for each document, the signed score differences $\Delta_{f-n} = s_f - s_n$ and $\Delta_{m-n} = s_m - s_n$, so points at the origin indicate no effect of gender substitution. Three patterns are consistent across models. First, a cluster around the origin shows that many documents are scored similarly across variants. Second, the distribution is skewed toward the lower-left quadrant: gendered variants tend to score lower than neutral ones. Third, $\Delta_{f-n}$ and $\Delta_{m-n}$ are correlated but not identical, with their correlation ranging from 0.83 to 0.89 across models, and most points above the $y=x$ diagonal, indicating that male variants tend to score higher than female variants.
These score asymmetries motivate our circuit analysis, which localizes the components responsible.

\section{Full Patching Results} \label{appendix:full_patching_results}

\subsection{Activation patching results} \label{appendix:full_activation_patching_results}

\begin{figure*}[h!]
    \centering
    \includegraphics[width=\linewidth]{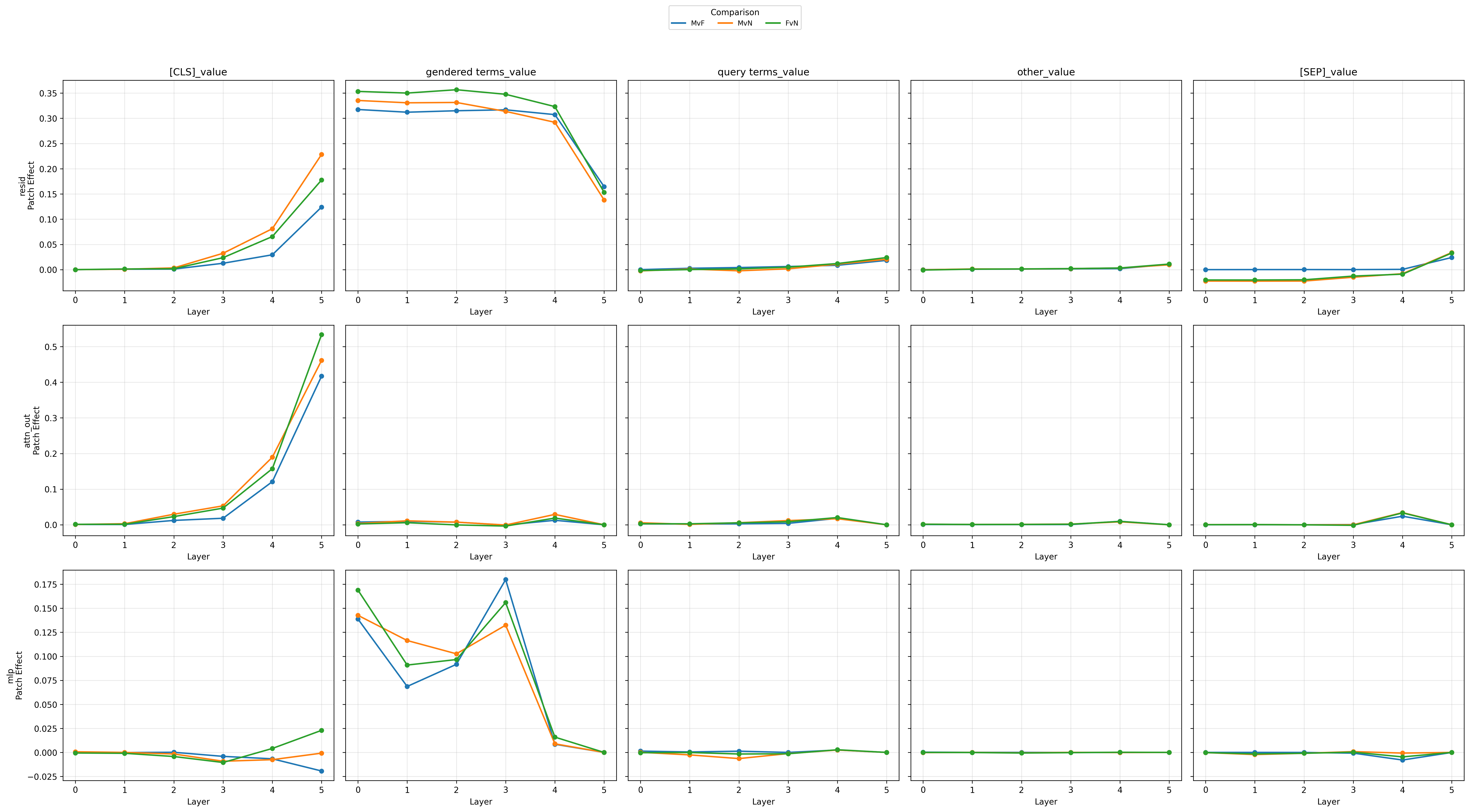}
    \caption{Activation patching results for TAS-B$_\text{ms-marco}$.}
    \label{fig:tasb_act_patch}
\end{figure*}

\begin{figure*}[h!]
    \centering
    \includegraphics[width=\linewidth]{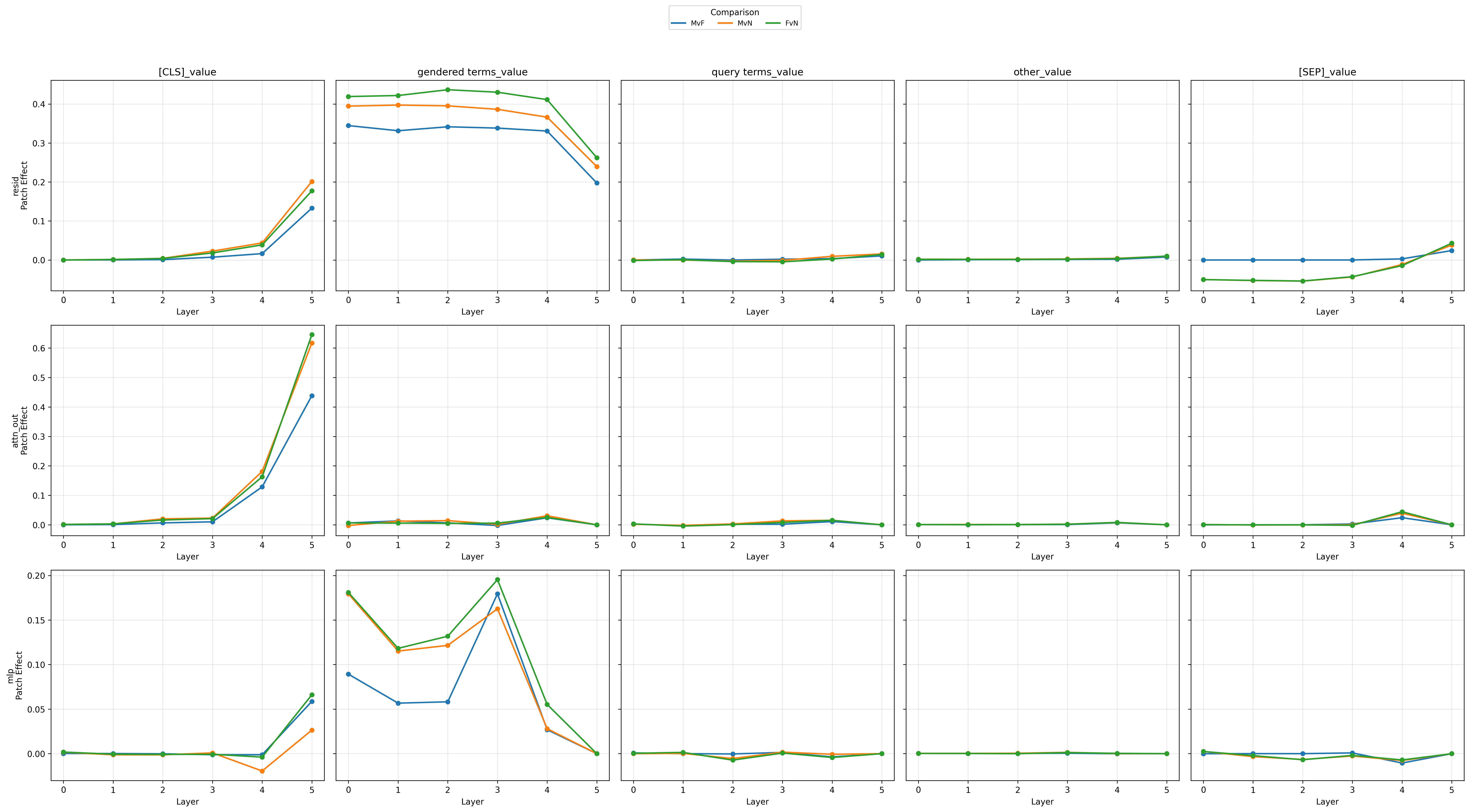}
    \caption{Activation patching results for DistilBERT$_\text{multi-qa}$.}
    \label{fig:distilbert_multiqa_act_patch}
\end{figure*}

\begin{figure*}[h!]
    \centering
    \includegraphics[width=\linewidth]{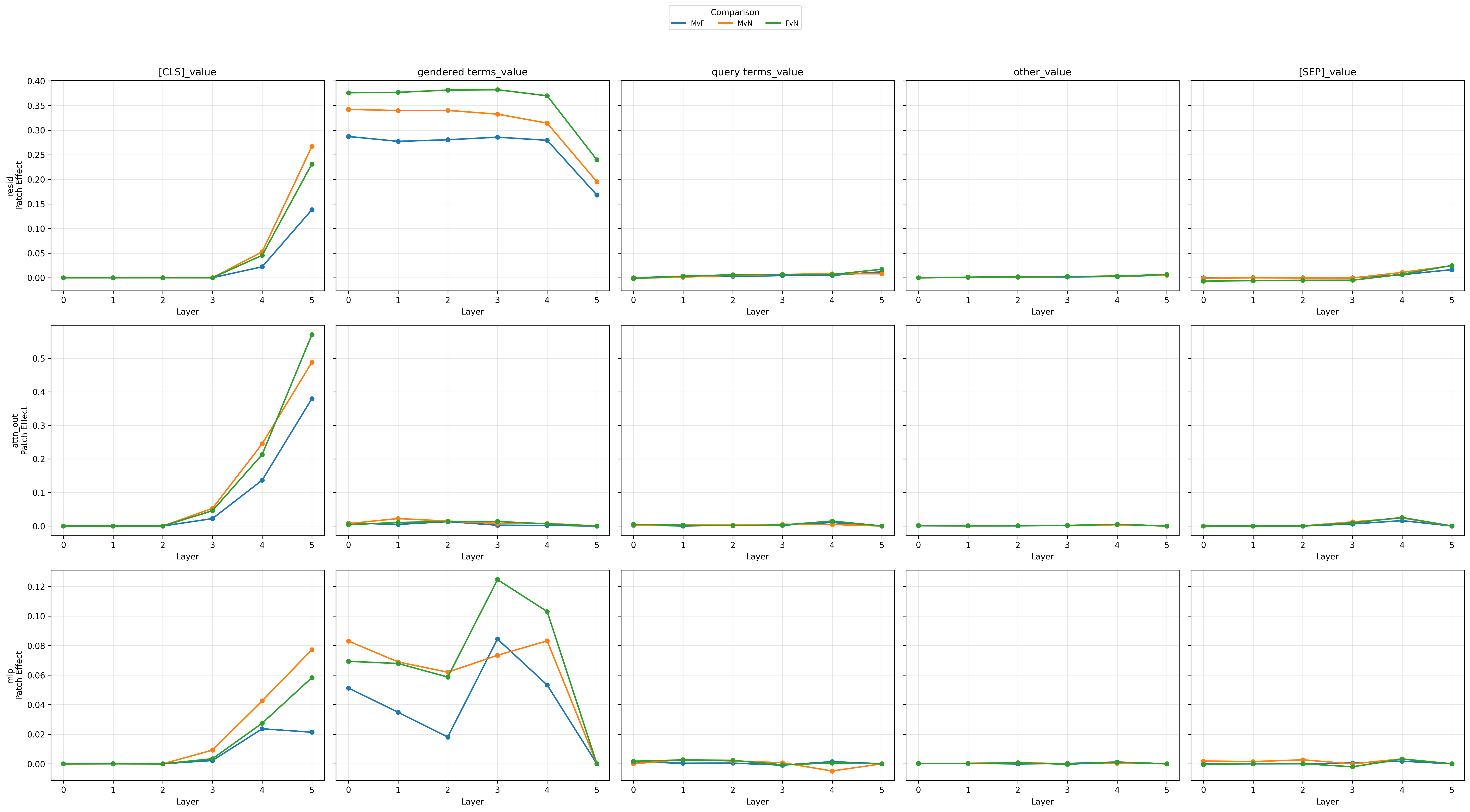}
    \caption{Activation patching results for MiniLM$_\text{multi-qa}$.}
    \label{fig:multiqa_minilm_act_patch}
\end{figure*}

\begin{figure*}[h!]
    \centering
    \includegraphics[width=\linewidth]{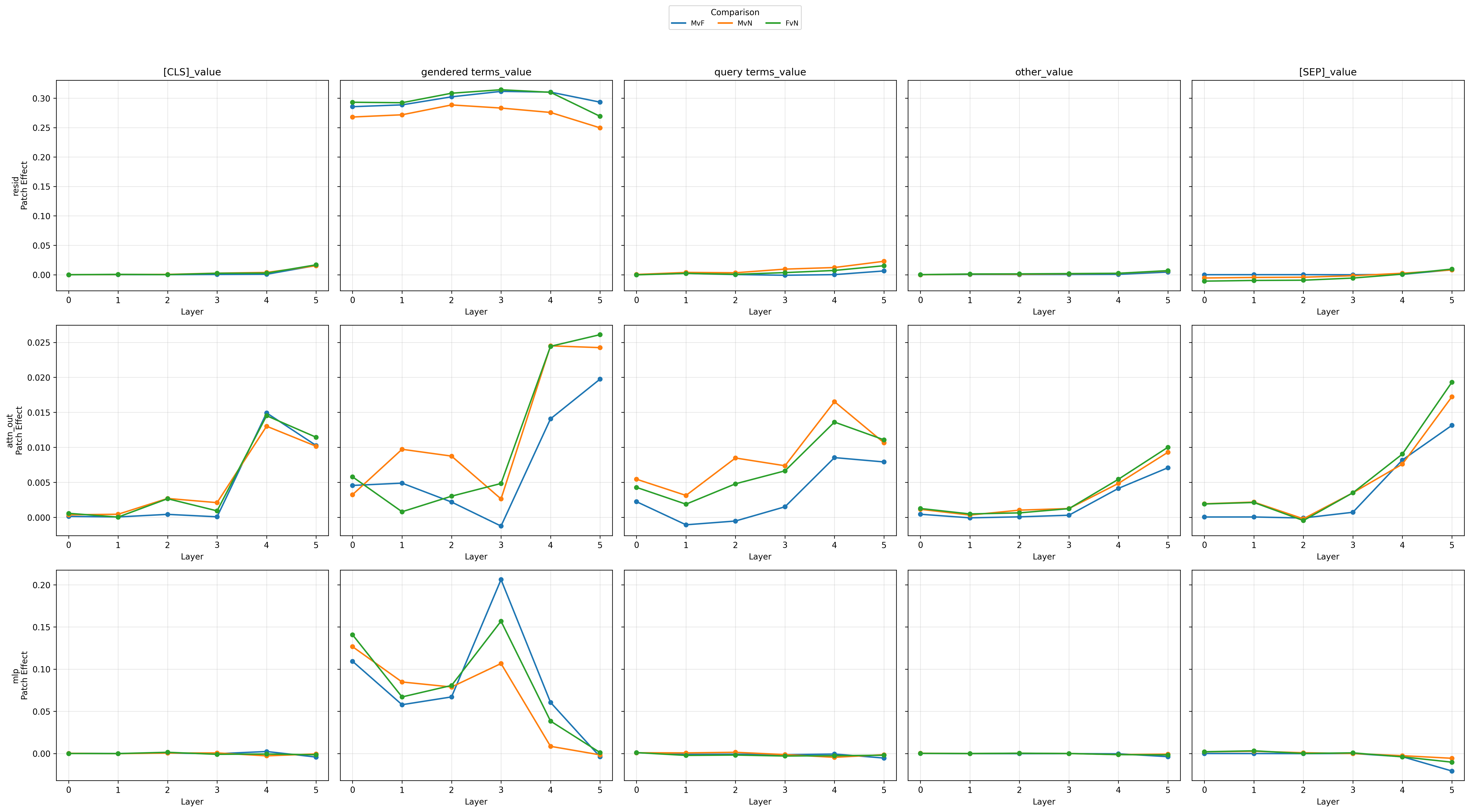}
    \caption{Activation patching results for DistilBERT$_\text{ms-marco}$.}
    \label{fig:msmarco_distilbert_act_patch}
\end{figure*}

\begin{figure*}[h!]
    \centering
    \includegraphics[width=\linewidth]{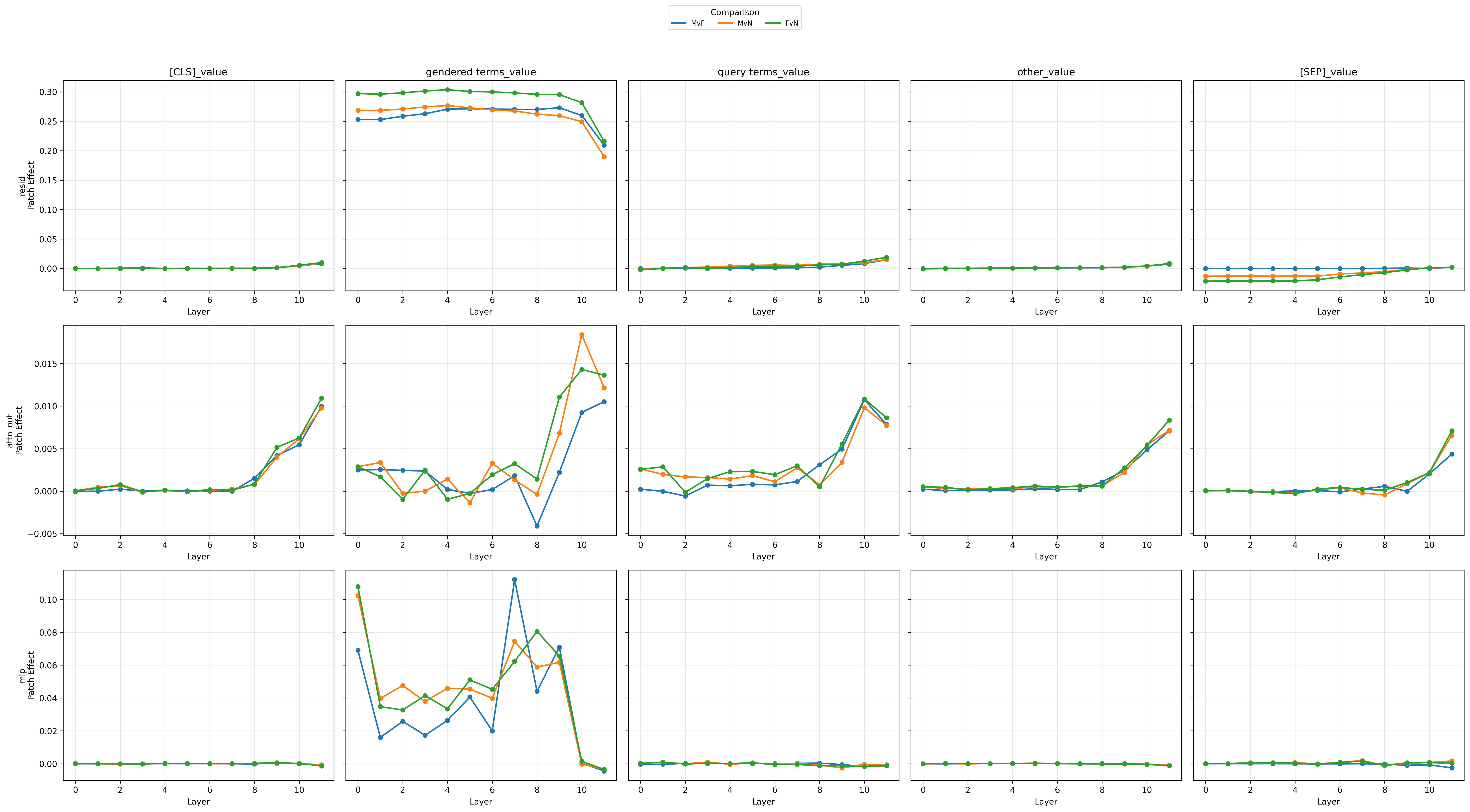}
    \caption{Activation patching results for BERT$_\text{ms-marco}$.}
    \label{fig:bert_act_patch}
\end{figure*}

\Cref{fig:bert_act_patch,fig:msmarco_distilbert_act_patch,fig:distilbert_multiqa_act_patch,fig:multiqa_minilm_act_patch,fig:tasb_act_patch} present activation patching results for all models and all token categories. In addition to the residual stream and attention block outputs, we also include results from inital experiments, which include patching MLP outputs. We leave further investigation of MLP effects to future work.


\subsection{Path patching results} \label{appendix:full_path_patching_results}

We consider components to be in the computational path if they are at least two standard deviations away from the mean patching effect across all heads for each comparison.

\paragraph{Patching to the final residual stream.}

\begin{figure*}
    \centering
    \includegraphics[width=\textwidth]{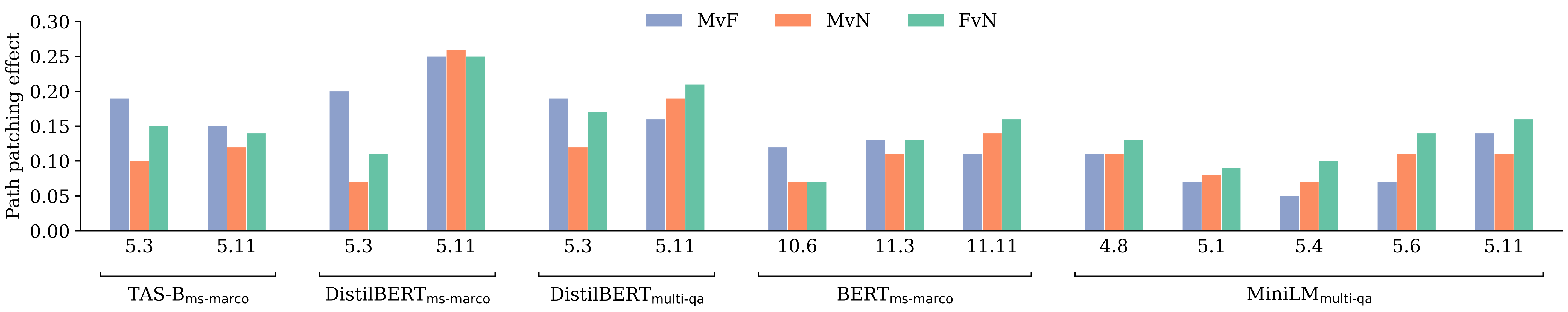}
    \caption{Path patching to the final residual stream for all models.}
    \label{fig:path_path_to_final_resid_all_models}
\end{figure*}

Figure~\ref{fig:path_path_to_final_resid_all_models} shows the results of path patching to the final residual stream for all models.

\paragraph{Patching to upstream attention heads.}

Table~\ref{tab:pp_to_upstream} shows the results of path patching from upstream attention heads to the heads identified from patching to the final residual stream.

\begin{table*}[h!]
\centering
\small
\setlength{\tabcolsep}{5pt}
\begin{tabular}{lllccc}
\toprule
Model & Receiver & Sender & MvF & MvN & FvN \\
\midrule
\multirow{6}{*}{DistilBERT-TAS-B$_{\text{ms-marco}}$}
& \multirow{2}{*}{5.3}
& 4.1  & 0.08 & 0.03 & 0.04 \\
&      & 4.10 & 0.06 & 0.02 & 0.03 \\
\cmidrule(lr){2-6}
& \multirow{3}{*}{5.11}
& 4.1  & 0.03 & 0.02 & 0.02 \\
&      & 4.5  & 0.02 & 0.02 & 0.02 \\
&      & 4.10 & 0.02 & --   & --   \\
\midrule
\multirow{6}{*}{DistilBERT$_{\text{ms-marco}}$}
& \multirow{3}{*}{5.3}
&        4.1  & 0.02 & --   & --   \\
&      & 4.5  & 0.02 & --   & --   \\
&      & 4.10 & 0.02 & --   & --   \\
\cmidrule(lr){2-6}
& \multirow{3}{*}{5.11}
&        4.1  & --   & 0.03 & --   \\
&      & 4.5  & --   & 0.02 & 0.02 \\
&      & 4.10 & --   & --   & --   \\
\midrule
\multirow{6}{*}{DistilBERT$_{\text{multi-qa}}$}
& \multirow{3}{*}{5.3}
&        4.1  & 0.07 & 0.03 & 0.04 \\
&      & 4.5  & --   & --   & 0.02 \\
&      & 4.10 & 0.03 & --   & --   \\
\cmidrule(lr){2-6}
& \multirow{3}{*}{5.11}
&        4.1  & 0.03 & --   & -- \\
&      & 4.5  & --   & 0.04 & 0.04 \\
\midrule
\multirow{2}{*}{BERT$_{\text{ms-marco}}$}
& 11.3  & 10.6 & 0.04 & --   & --   \\
& 11.11 & 10.6 & 0.03 & 0.02 & 0.03 \\
\midrule
MiniLM$_{\text{multi-qa}}$
& 5.11 & 4.3 & 0.05 & 0.02 & 0.04 \\
\bottomrule
\end{tabular}
\caption{Path patching results \emph{to downstream attention heads} (Receiver, identified from path patching to the final residual stream) from upstream heads (Sender), with MLPS recomputed. Only heads with patching effects greater than 2SD from the mean are reported.}
\label{tab:pp_to_upstream}
\end{table*}

\section{Full Ablation Results}

\begin{figure*}[h!]
    \centering
    \includegraphics[width=0.95\linewidth]{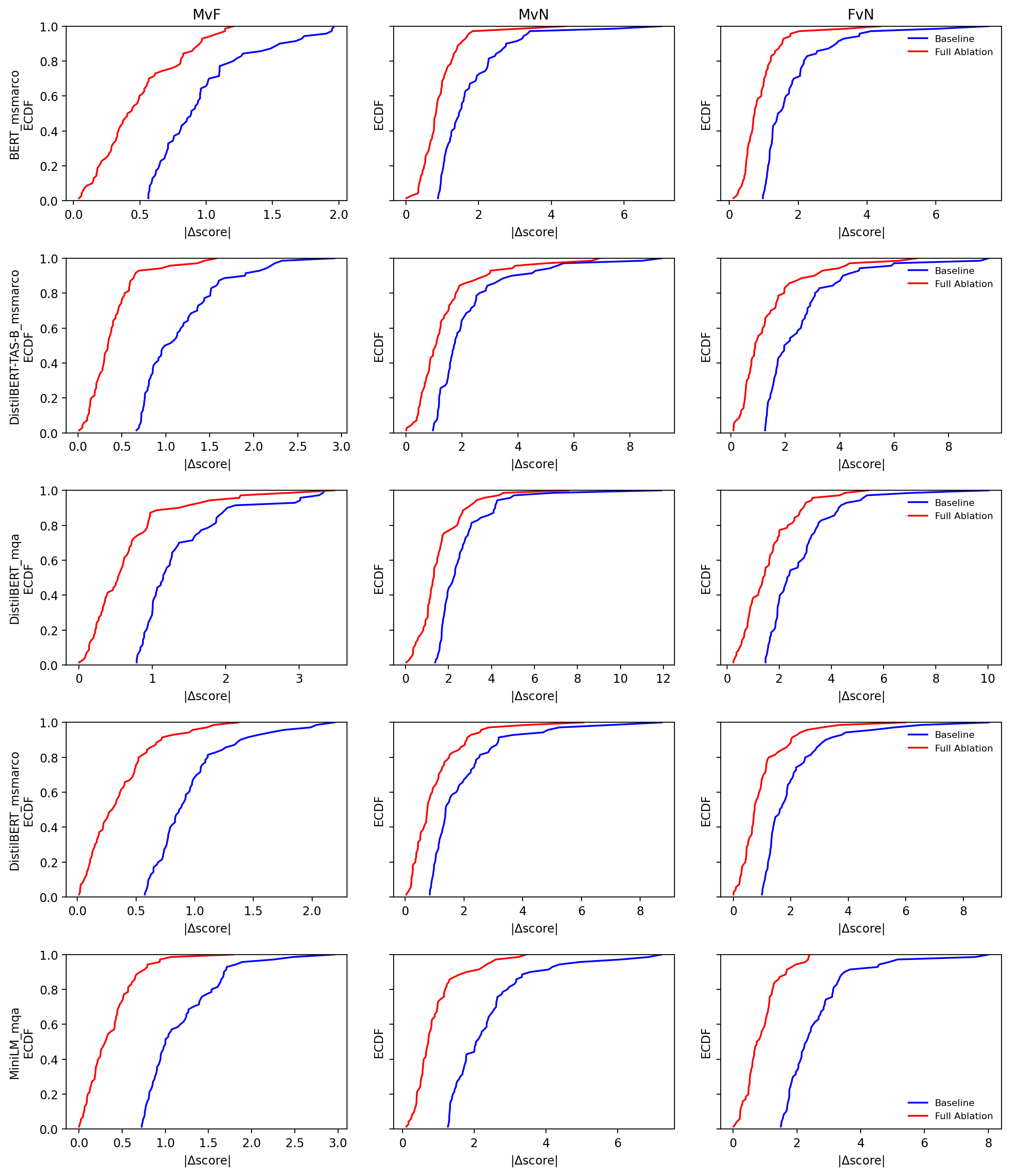}
    \caption{Mean ablating identified heads from path patching; top 30 percent of score differences.}
    \label{fig:ecdf_ablation}
\end{figure*}

To confirm the causal role of the heads identified by path patching, we mean ablate each identified head and measure the resulting reduction in score differences between contrastive pairs. We replace the head's output with its mean activation, computed per head across all token positions and all documents in the dataset. Figure~\ref{fig:ecdf_ablation} shows the mean ablation results, zoomed into the top 30\% of score differences within each gender comparison for better visualization of effects.

\begin{figure*}[h!]
    \centering
    \includegraphics[width=0.95\linewidth]{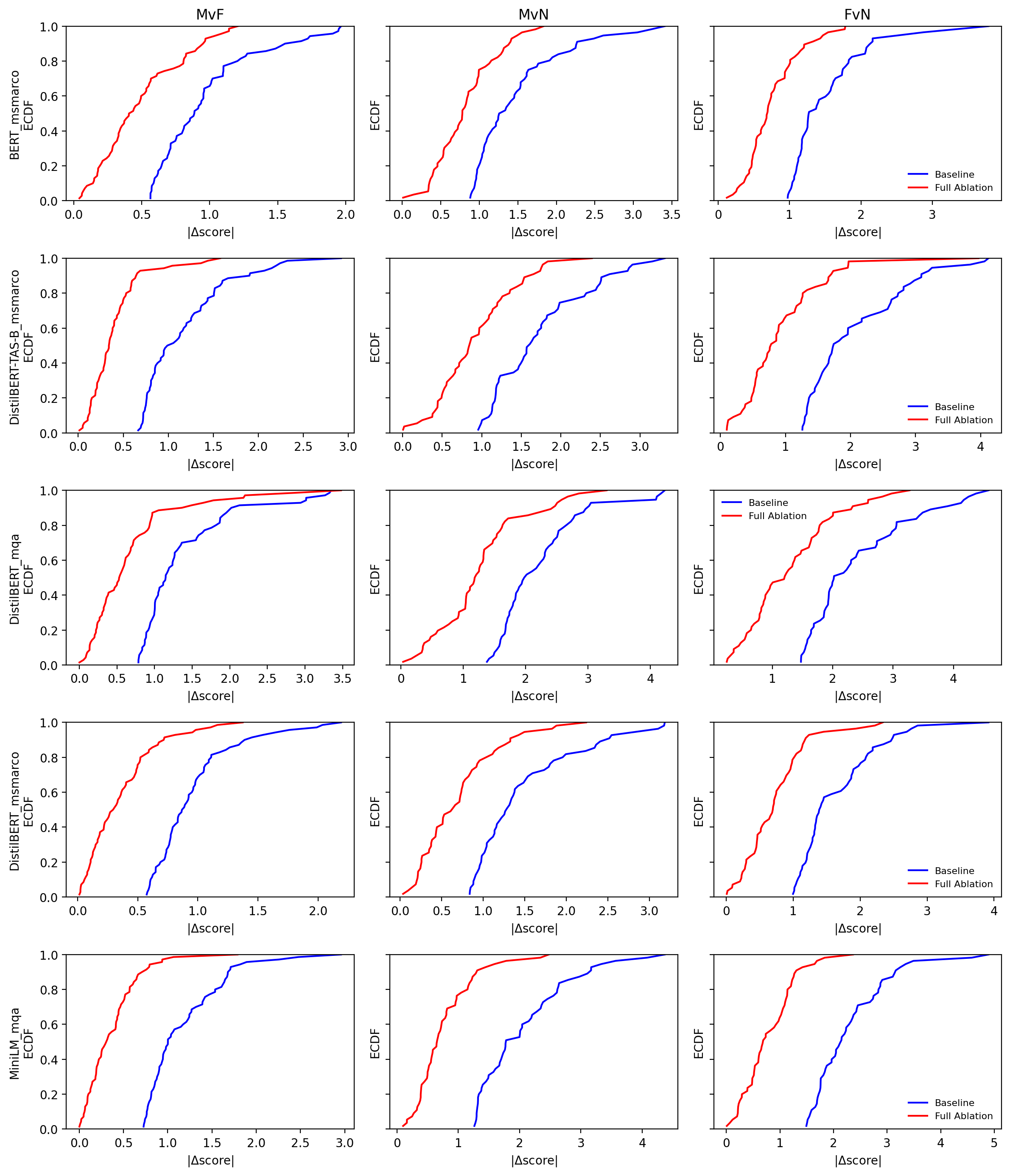}
    \caption{Mean ablation results filtering out documents where the neutral variant's gendered term appears in the query; top 30 percent of score differences.}
    \label{fig:mean_ablation_filter_neutral}
\end{figure*}

In Section~\ref{sec:results_path}, we note that for the MvN and FvN comparisons, ablation does not reduce score differences for the pairs with the largest score differences. Section~\ref{sec:results_behavior} attributes these pairs to documents where the neutral term also appears in the query. Figure~\ref{fig:mean_ablation_filter_neutral} shows the mean ablation results after filtering out these documents, in which the effect of ablation is more uniform across the remaining samples.

\section{Full Steering Results} \label{appendix:steering}

This section provides additional details on the steering experiments discussed in Section~\ref{sec:steering}.

\subsection{Implementation details}
We compute steering vectors for all three pairwise gender comparisons (MvF, MvN, and FvN). For each comparison, the steering vector is the mean difference in activations between the two variants, computed over the relevant document pairs. At the embedding level, steering is applied at the token level, modifying only the gendered token's embedding. At the attention level, steering is applied across all token positions at the identified final-layer attention heads.

\subsection{Results across all models}

\begin{figure*}[h!]
    \centering
    \includegraphics[width=0.95\linewidth]{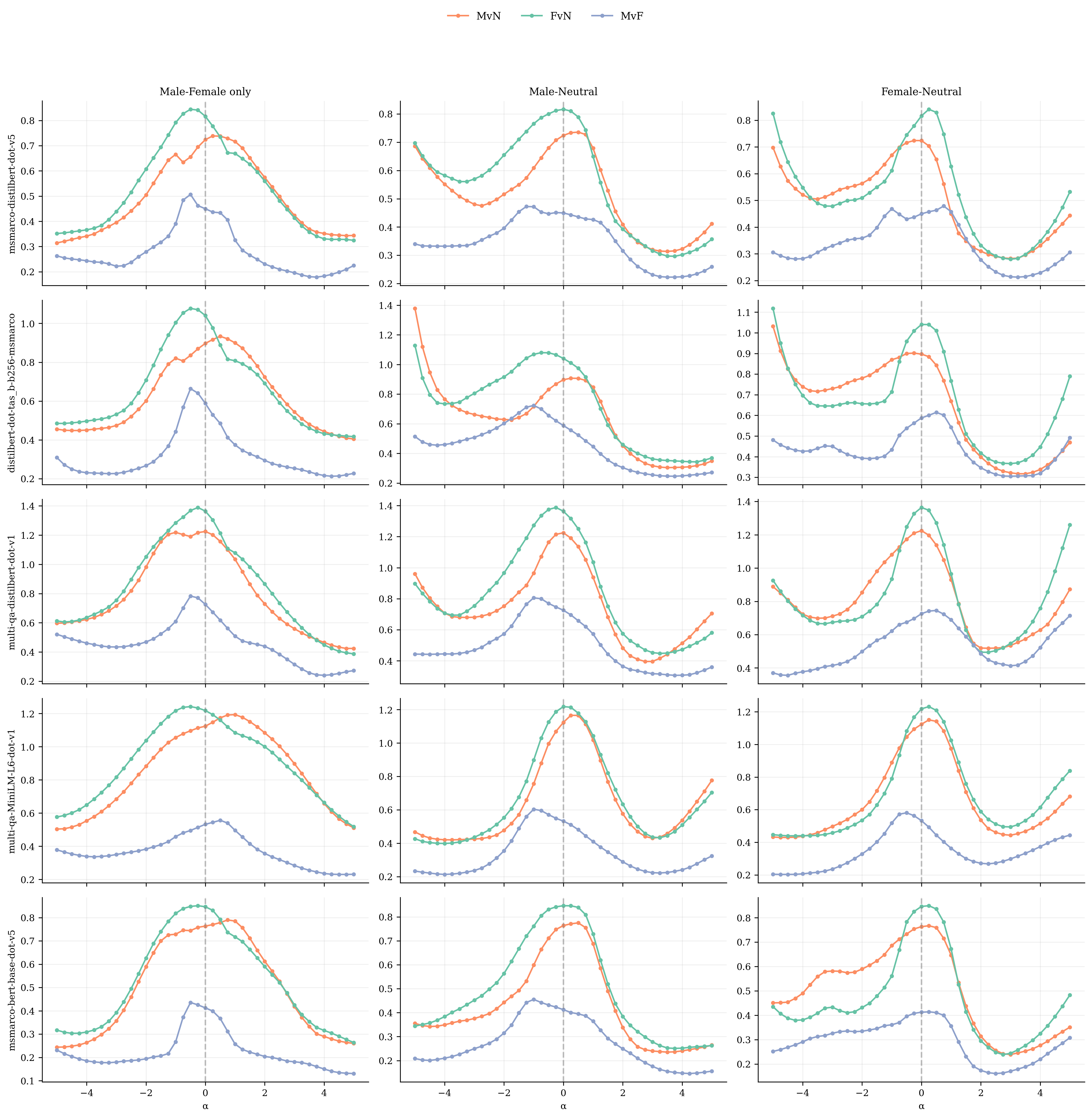}
    \caption{Embedding steering results for all five models for all comparison directions, with $\alpha \in [-5,5]$ in 0.25 increments.}
    \label{fig:steering_embed_all_models}
\end{figure*}

\begin{figure*}[h!]
    \centering
    \includegraphics[width=0.95\linewidth]{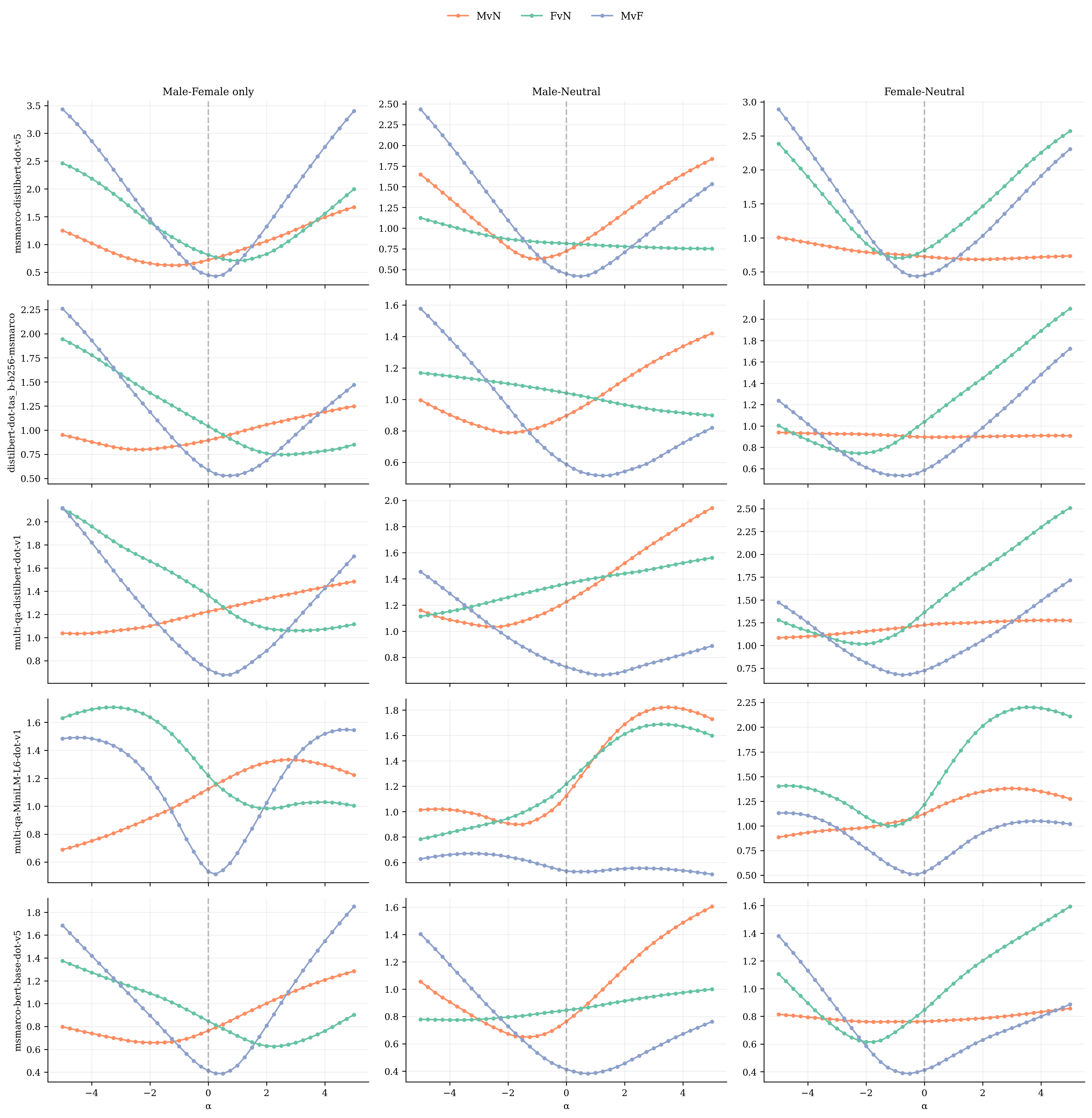}
    \caption{Head steering results for all five models for all comparison directions, with $\alpha \in [-5,5]$ in 0.25 increments.}
    \label{fig:steering_head_all_models}
\end{figure*}

\Cref{fig:steering_embed_all_models,fig:steering_head_all_models} present steering results for all five models at both intervention points (embedding-level and attention-level).

\begin{figure*}[h!]
    \centering
    \includegraphics[width=0.6\linewidth]{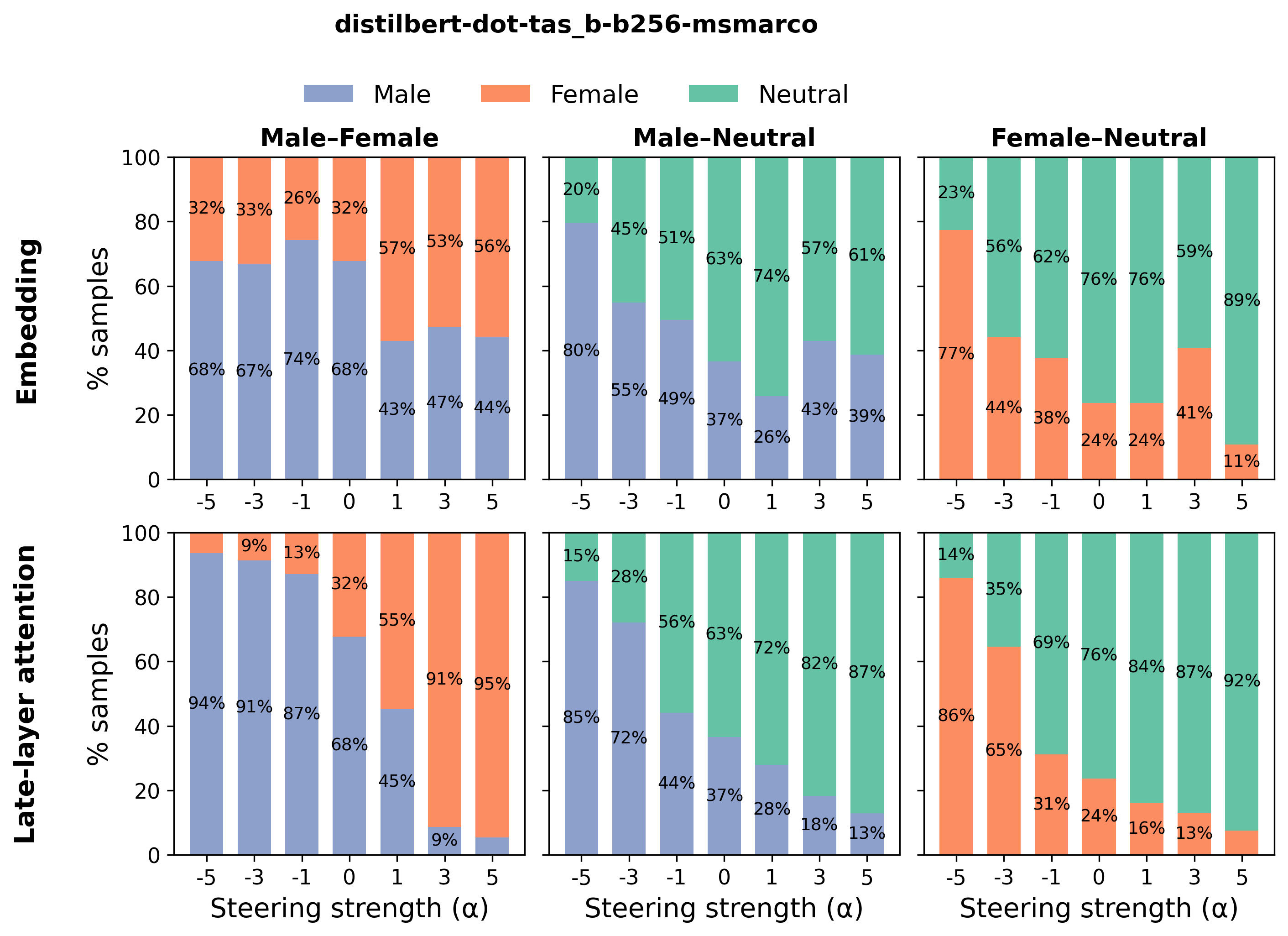}
    \caption{Steering effects on the embeddings (top) compared to attention (bottom) for TAS-B$_\text{ms-marco}$, with $\alpha \in [-5,5]$ in increments of 2 compared to the baseline ($\alpha = 0$). Percentages refer to the proportion of documents for the given gender that score higher than the other.}
    \label{fig:steering_bars_tasb}
\end{figure*}

\begin{figure*}[h!]
    \centering
    \includegraphics[width=0.6\linewidth]{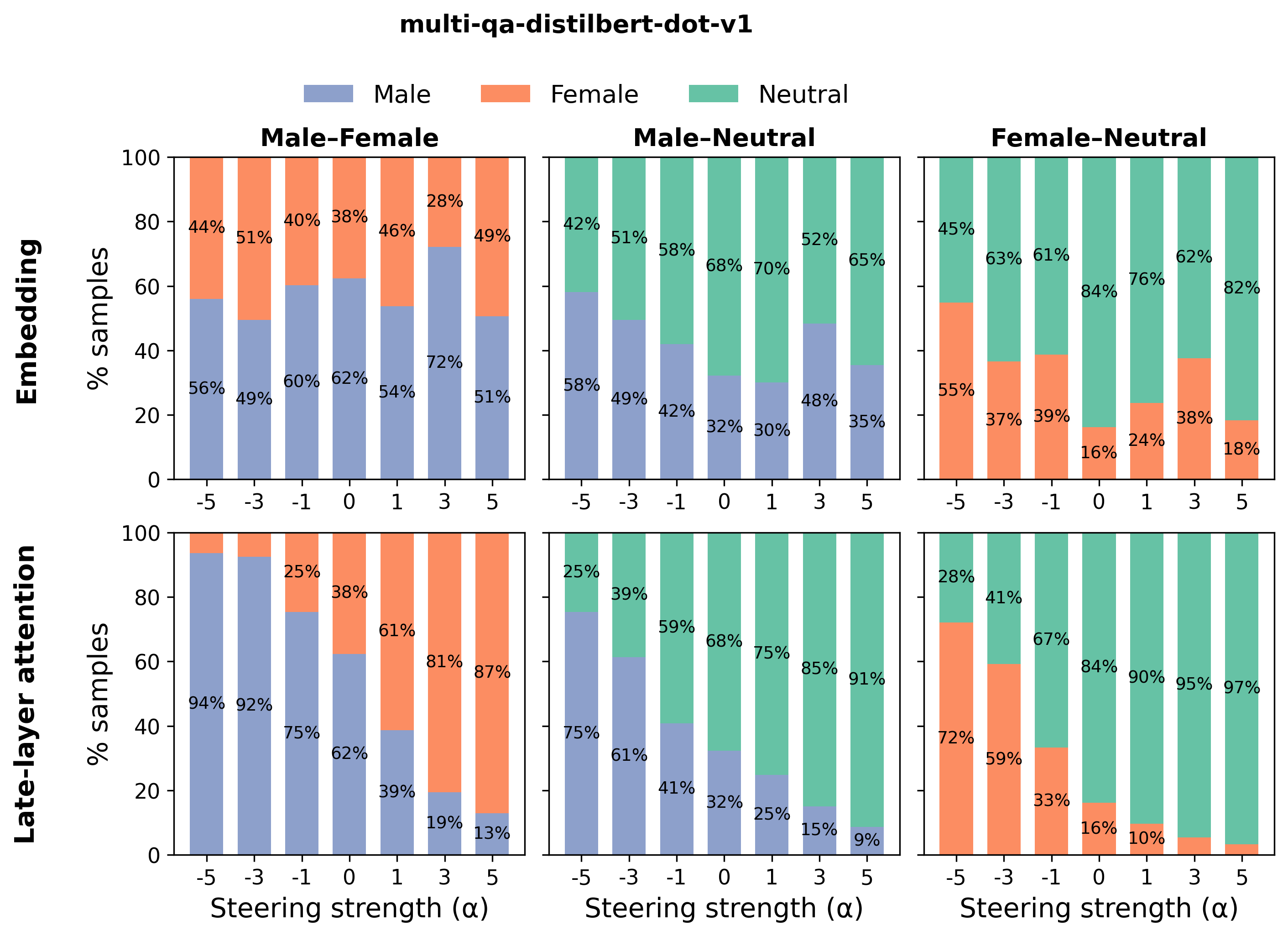}
    \caption{Steering effects on the embeddings (top) compared to attention (bottom) for DistilBERT$_\text{multi-qa}$.}
    \label{fig:steering_bars_multiqa_distilbert}
\end{figure*}

\begin{figure*}[h!]
    \centering
    \includegraphics[width=0.6\linewidth]{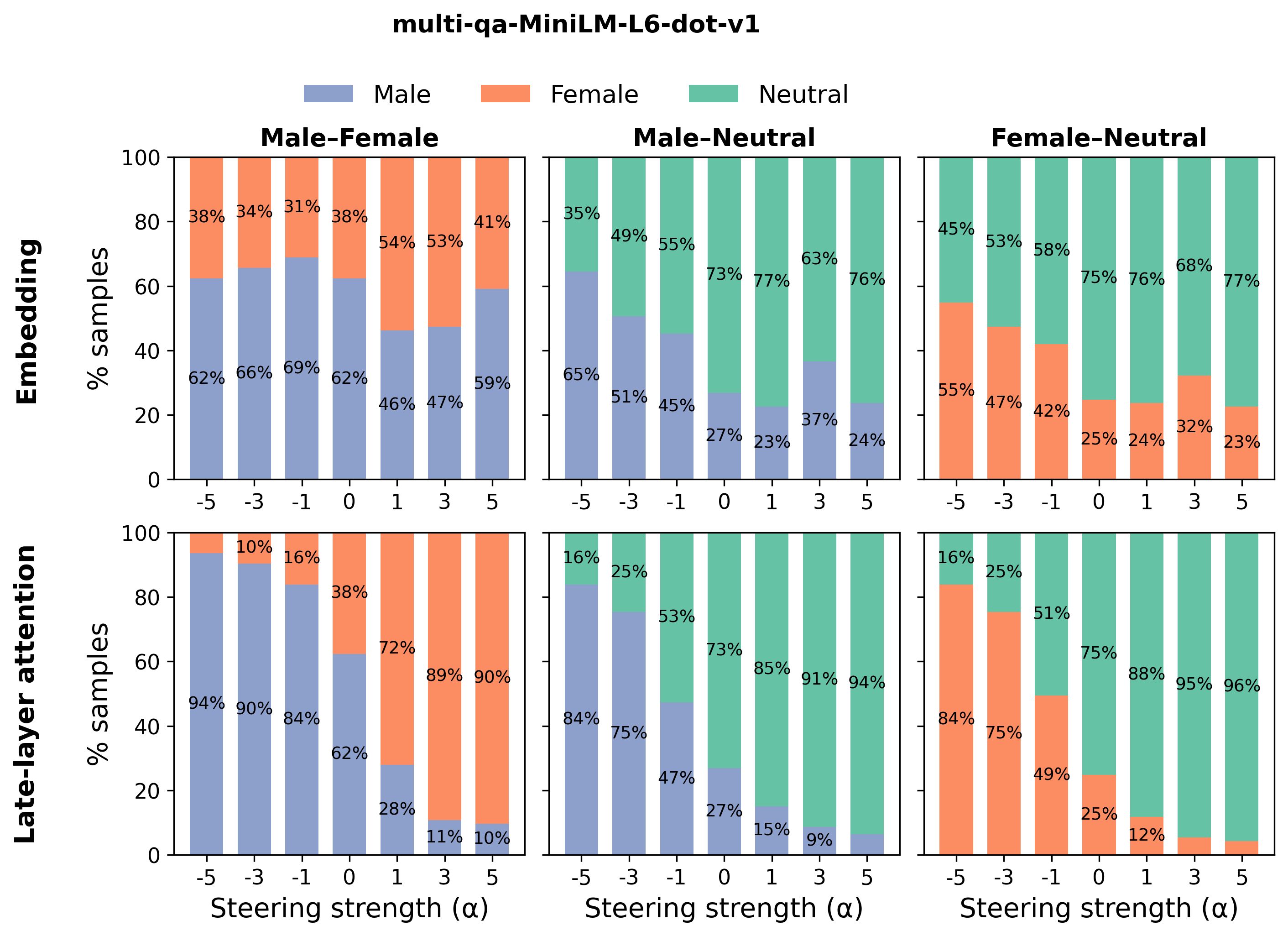}
    \caption{Steering effects on the embeddings (top) compared to attention (bottom) for Mini-LM$_\text{multi-qa}$.}
    \label{fig:steering_bars_multiqa_minilm}
\end{figure*}

\begin{figure*}[h!]
    \centering
    \includegraphics[width=0.6\linewidth]{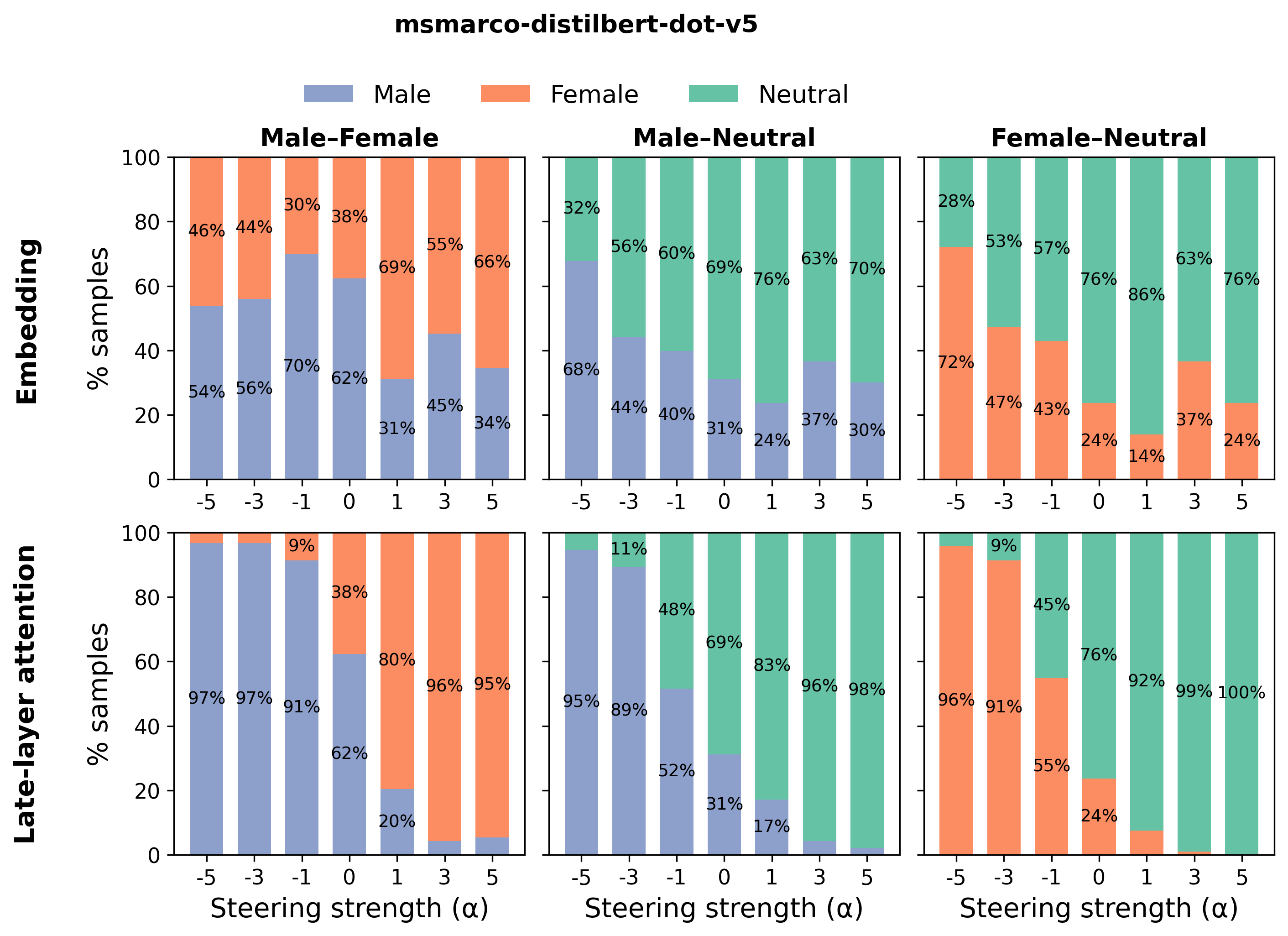}
    \caption{Steering effects on the embeddings (top) compared to attention (bottom) for DistilBERT$_\text{ms-marco}$.}
    \label{fig:steering_bars_msmarco_distilbert}
\end{figure*}

\begin{figure*}[h!]
    \centering
    \includegraphics[width=0.6\linewidth]{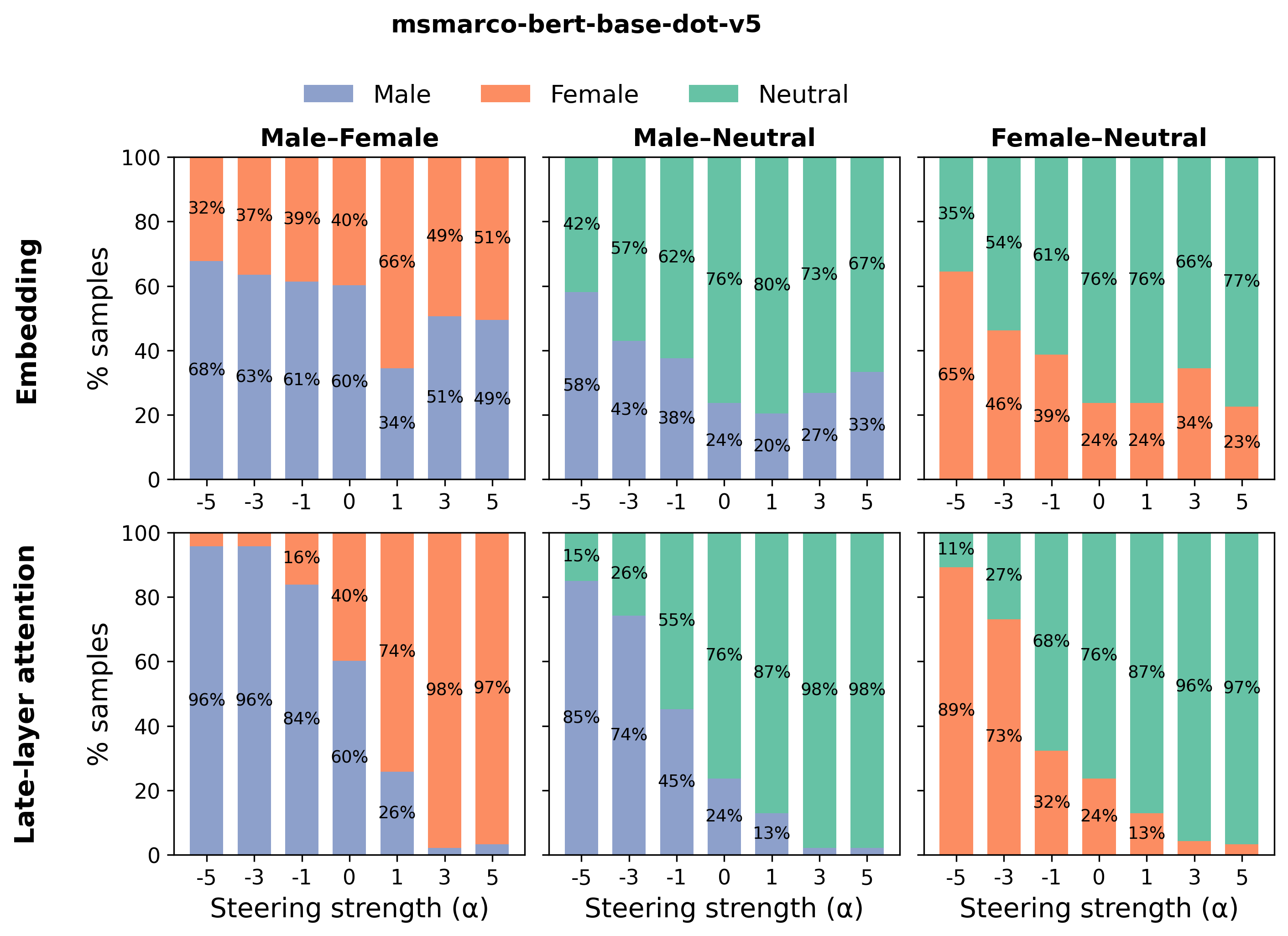}
    \caption{Steering effects on the embeddings (top) compared to attention (bottom) for BERT$_\text{ms-marco}$.}
    \label{fig:steering_bars_msmarco_bert}
\end{figure*}

In addition to score difference visualizations, we report \textit{preference rates}, which we define as the proportion of document pairs in which one variant scores higher than the other, for the steering coefficients (\Cref{fig:steering_bars_msmarco_bert,fig:steering_bars_msmarco_distilbert,fig:steering_bars_tasb,fig:steering_bars_multiqa_distilbert,fig:steering_bars_multiqa_minilm}). These preference rates provide a complementary view to score difference changes and whether steering systematically ``flips'' the preference between variants.



\end{document}